\documentstyle[12pt,epsf]{article}

\catcode`\@=11
\long\def\@makefntext#1{
\protect\noindent \hbox to 3.2pt {\hskip-.9pt  
$^{{\ninerm\@thefnmark}}$\hfil}#1\hfill}                

\def\@makefnmark{\hbox to 0pt{$^{\@thefnmark}$\hss}}  
        
\def\ps@myheadings{\let\@mkboth\@gobbletwo
\def\@oddhead{\hbox{}
\rightmark\hfil\ninerm\thepage}   
\def\@oddfoot{}\def\@evenhead{\ninerm\thepage\hfil
\leftmark\hbox{}}\def\@evenfoot{}
\def\sectionmark##1{}\def\subsectionmark##1{}}

\renewcommand{\thefootnote}{\fnsymbol{footnote}}

\newcounter{sectionc}\newcounter{subsectionc}\newcounter{subsubsectionc}
\renewcommand{\section}[1] {\vspace*{0.6cm}\addtocounter{sectionc}{1} 
\setcounter{subsectionc}{0}\setcounter{subsubsectionc}{0}\noindent 
        {\normalsize\bf\thesectionc. #1}\par\vspace*{0.4cm}}
\renewcommand{\subsection}[1] {\vspace*{0.6cm}\addtocounter{subsectionc}{1} 
        \setcounter{subsubsectionc}{0}\noindent 
        {\normalsize\it\thesectionc.\thesubsectionc. #1}\par\vspace*{0.4cm}}
\renewcommand{\subsubsection}[1]
{\vspace*{0.6cm}\addtocounter{subsubsectionc}{1}
        \noindent {\normalsize\rm\thesectionc.\thesubsectionc.\thesubsubsectionc. 
        #1}\par\vspace*{0.4cm}}

\newcounter{appendixc}
\newcounter{subappendixc}[appendixc]
\newcounter{subsubappendixc}[subappendixc]

\renewcommand{\appendix}[1] {\vspace*{0.6cm}
        \refstepcounter{appendixc}
        \setcounter{figure}{0}
        \setcounter{table}{0}
        \setcounter{equation}{0}
        \renewcommand{\thefigure}{\Alph{appendixc}.\arabic{figure}}
        \renewcommand{\thetable}{\Alph{appendixc}.\arabic{table}}
        \renewcommand{\theappendixc}{\Alph{appendixc}}
        \renewcommand{\theequation}{\Alph{appendixc}.\arabic{equation}}
        \noindent{\bf Appendix \theappendixc #1}\par\vspace*{0.4cm}}

\def\abstracts#1{{
        \centering{\begin{minipage}{12.2truecm}\footnotesize\baselineskip=12pt\noindent
        \centerline{\footnotesize ABSTRACT}\vspace*{0.3cm}
        \parindent=0pt #1
        \end{minipage}}\par}} 

\newcounter{itemlistc}
\newcounter{romanlistc}
\newcounter{alphlistc}
\newcounter{arabiclistc}

\newcommand{\fcaption}[1]{
        \refstepcounter{figure}
        \setbox\@tempboxa = \hbox{\footnotesize Fig.~\thefigure. #1}
        \ifdim \wd\@tempboxa > 6in
           {\begin{center}
        \parbox{6in}{\footnotesize\baselineskip=12pt Fig.~\thefigure. #1}
            \end{center}}
        \else
             {\begin{center}
             {\footnotesize Fig.~\thefigure. #1}
              \end{center}}
        \fi}

\newcommand{\tcaption}[1]{
        \refstepcounter{table}
        \setbox\@tempboxa = \hbox{\footnotesize Table~\thetable. #1}
        \ifdim \wd\@tempboxa > 6in
           {\begin{center}
        \parbox{6in}{\footnotesize\baselineskip=12pt Table~\thetable. #1}
            \end{center}}
        \else
             {\begin{center}
             {\footnotesize Table~\thetable. #1}
              \end{center}}
        \fi}

\def\@citex[#1]#2{\if@filesw\immediate\write\@auxout
        {\string\citation{#2}}\fi
\def\@citea{}\@cite{\@for\@citeb:=#2\do
        {\@citea\def\@citea{,}\@ifundefined
        {b@\@citeb}{{\bf ?}\@warning
        {Citation `\@citeb' on page \thepage \space undefined}}
        {\csname b@\@citeb\endcsname}}}{#1}}

\newif\if@cghi
\def\cite{\@cghitrue\@ifnextchar [{\@tempswatrue
        \@citex}{\@tempswafalse\@citex[]}}
\def\citelow{\@cghifalse\@ifnextchar [{\@tempswatrue
        \@citex}{\@tempswafalse\@citex[]}}
\def\@cite#1#2{{$\null^{#1}$\if@tempswa\typeout
        {IJCGA warning: optional citation argument 
        ignored: `#2'} \fi}}

 1
 1
 1

\font\ninerm=cmr9

\newcommand{\lesssim}{\mathrel{\rlap{\lower4pt\hbox{\hskip1pt$\sim$}}
    \raise1pt\hbox{$<$}}}         
\newcommand{\gtrsim}{\mathrel{\rlap{\lower4pt\hbox{\hskip1pt$\sim$}}
    \raise1pt\hbox{$>$}}}         

\begin{document}
\thispagestyle{empty}

\rightline{\normalsize\rm MPI-PhT/98-63}
\rightline{\normalsize\rm August 1998}

\vspace*{2cm}

\centerline{\large\bf Evidence for ``sterile neutrino'' dark matter?}

\vspace*{1cm}

\centerline{\footnotesize Paolo Gondolo}
\baselineskip=13pt
\centerline{\footnotesize\it Max Planck Institut f\"ur Physik, 
F\"ohringer Ring 6}
\baselineskip=12pt
\centerline{\footnotesize\it  80805, Munich, Germany}
\centerline{\footnotesize E-mail: gondolo@mppmu.mpg.de}

\vspace*{1cm} 
\abstracts{ I show that it may be possible to explain the
  present evidence for a gamma-ray emission from the galactic halo as due to
  halo WIMP annihilations. Not only the intensity and spatial pattern of the
  halo emission can be matched but also the relic density of the candidate WIMP
  can be in the cosmologically interesting domain.  After a model-independent
  analysis to learn about the properties of a suitable candidate, I present a
  working model: a sterile neutrino in a model with an extended Higgs sector.
  Two examples indicate the existence of an interesting region in the model
  parameter space where present observational and experimental constraints
  are satisfied and the gamma-ray emission is reproduced.  }
 
\vspace*{3cm}

\begin{center}
{\it Talk presented at the Ringberg Euroconference ``New Trends in Neutrino
  Physics,'' Ringberg Castle, Tegernsee, Germany, 24--29 May 1998.}
\end{center}

\newpage
\setcounter{page}{1}

\centerline{\normalsize\bf Evidence for ``sterile neutrino'' dark matter?}

\centerline{\footnotesize Paolo Gondolo}
\baselineskip=13pt
\centerline{\footnotesize\it Max Planck Institut f\"ur Physik, 
F\"ohringer Ring 6}
\baselineskip=12pt
\centerline{\footnotesize\it  80805, Munich, Germany}
\centerline{\footnotesize E-mail: gondolo@mppmu.mpg.de}

\vspace*{0.9cm} 
\abstracts{ I show that it may be possible to explain the
  present evidence for a gamma-ray emission from the galactic halo as due to
  halo WIMP annihilations. Not only the intensity and spatial pattern of the
  halo emission can be matched but also the relic density of the candidate WIMP
  can be in the cosmologically interesting domain.  After a model-independent
  analysis to learn about the properties of a suitable candidate, I present a
  working model: a sterile neutrino in a model with an extended Higgs sector.
  Two examples indicate the existence of an interesting region in the model
  parameter space where present observational and experimental constraints
  are satisfied and the gamma-ray emission is reproduced.  }

\normalsize\baselineskip=15pt
\setcounter{footnote}{0}
\renewcommand{\thefootnote}{\alph{footnote}}

\section{Introduction}

A sophisticated analysis of EGRET data (Dixon et al.\ 1998) has found evidence
for gamma-ray emission from the galactic halo. Filtering the data with a
wavelet expansion, Dixon et al.\ have subtracted an
isotropic extra-galactic component and expected contributions from cosmic ray
interactions with the interstellar gas and from inverse Compton of ambient
photons by cosmic ray electrons, and they have produced a map of the intensity
distribution of the residual gamma-ray emission. Besides a few ``point'' sources, they find an
excess in the central region extending somewhat North of the galactic plane,
and a weaker emission from regions in the galactic halo. They mention an
astrophysical interpretation for this halo emission: inverse Compton by cosmic
ray electrons distributed on larger scales than those commonly discussed and
with anomalously hard energy spectrum.

I find it intriguing that the angular distribution of the halo emission
resembles that expected from pair annihilation of dark matter WIMPs in the
galactic halo (Gunn et al.\ 1978, Turner 1986), and moreover, that the
gamma-ray intensity is similar to that expected from annihilations of a thermal
relic with present mass density 0.1--0.2 of the critical density (Gondolo
1998a). Namely, the emission at $ b \ge 20^\circ $ is approximately constant at
a given angular distance from the galactic center -- except for a region around
$(b,l) = (60^\circ,45^\circ) $, correlated to the position of the Moon, and a
region around $(b,l) = (190^\circ, -30^\circ) $, where there is a local cloud
($b$ and $l$ are the galactic latitude and longitude, respectively). Dixon et
al.~(1998) argue against the possibility of WIMP annihilations on the base that
direct annihilation of neutralinos into photons would give too low a gamma-ray
signal.  However most of the photons from WIMP annihilations are usually not
produced directly but come from the decay of neutral pions generated in the
particle cascades following annihilation.

Presently preferred dark matter candidates tend to give a gamma emission which
is too low even in the continuum.\footnote{Attempts to increase the flux by
  clumpiness in the halo (Berstr\"om, Edsj\"o and Ullio 1998) tend to produce
  either too many antiprotons or too few photons below 1 GeV, in contrast to
  the Dixon et al.\ maps (see section 4).} 
In a search for a suitable candidate it is worth
examining the impact of various constraints on the properties of the candidate.
I will later introduce a working model, so to make the discussion more
concrete.

\section{A candidate WIMP can be a thermal relic.}

It is useful to use the WIMP mass $m_\chi$ and its annihilation cross section
(times relative velocity at $v=0$) $\sigma v$ as parameters in the discussion.
The requirement to approximately match the Dixon et al.\ maps selects a band in
the $\sigma v$--$m_\chi$ plane (band marked ``halo $\gamma$'s'' in fig.~1). Another
band is selected by the requirement that the WIMP is a thermal relic from the
early universe, with a relic abundance in the range $ 0.025 < \Omega h^2 < 1$
(band marked ``$ 0.025 < \Omega h^2 < 1$'' in fig.~1). The intersection of the
two bands defines the interesting region.

\begin{figure}[htb]
\centerline{\epsfxsize = 0.8\hsize \epsfbox{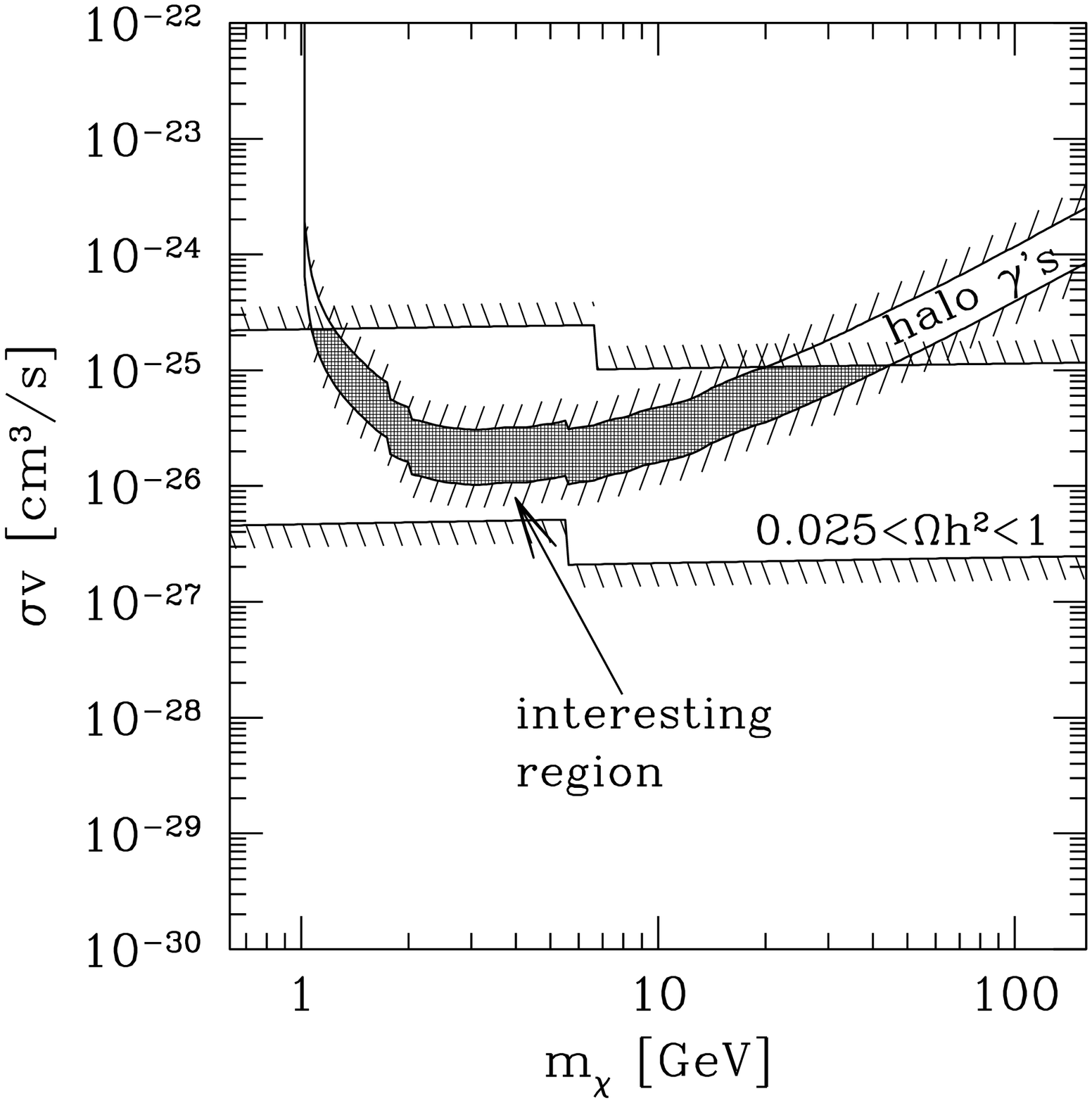}}
\fcaption{Region in the annihilation cross section versus mass
  plane where the gamma-ray halo may be explained by annihilations of a thermal
  dark matter relic.}
\label{fig:1}
\end{figure}

The gamma-ray band is obtained as follows.
The gamma-ray intensity 
from WIMP annihilations in the galactic halo from
direction with galactic longitude $b$ and galactic latitude $l$ is given by
\begin{equation}
\label{phigamma}
\phi_{\gamma}(b,l,\mathord{>}E) = n_{\gamma}(\mathord{>}E) \,
\frac{ \sigma v }{ 4 \pi m_\chi^2 } \, \int \rho_\chi^2 d l .
\end{equation}
$ \phi_{\gamma}(b,l,\mathord{>}E) $ is in photons/(cm$^2$~s~sr),
$n_{\gamma}(\mathord{>}E)$ is the number of photons with energy above $E$
generated per WIMP annihilation, $ \sigma v$ is the WIMP annihilation cross
section times relative velocity, $m_\chi$ is the WIMP mass, and $ \rho_\chi $
is the WIMP mass density in the halo. The integral in eq.~(\ref{phigamma}) is
along the line of sight in direction $(b,l)$.  This integral gives the angular
dependence of the gamma-ray flux and depends on details of the dark halo model,
which are not well known. For a canonical halo model, $ \rho(r) = \rho_{\rm
  loc} ( r_c^2 + R^2 )/( r_c^2 + r^2 ) $, where $ \rho_{\rm loc} $ is the WIMP
mass density in the solar neighborhood, $r_c$ is the halo core radius, $R$ is
the distance of the Sun from the galactic center, and $r$ is the galactocentric
distance. In this case, the integral depends only on the angle $ \psi $ between
the direction of observation and the galactic center,
\begin{equation}
\label{phiint}
\int \rho_\chi^2 d l  = \rho_{\rm loc}^2 R 
  \frac{x}{2(x-c^2)^{3/2}} \, \left[
  \frac{\pi}{2} + \arctan\frac{c}{\sqrt{x-c^2}} +
  \frac{ c \sqrt{x-c^2} }{x} \right] .
\end{equation}
Here $x=1+(r_c/R)^2$ and $c=\cos\psi=\cos b \cos l$.  From dynamical studies
one finds $\rho_{\rm loc}$ = 0.3--0.5 GeV/cm$^3$, $R$ = 8--8.5 kpc, and $r_c$ =
2--8 kpc.  It is interesting to notice that the gamma-ray intensities at $ \psi
= 40^\circ$ and $ \psi = 60^\circ$ are approximately in the ratio 2:1 as on the
Dixon et al.\ maps.  Assuming WIMP annihilation into quark-antiquark and
lepton-antilepton pairs, fixing $n_{\gamma}(\mathord{>}1{\rm GeV})$ with the
Lund Monte-Carlo, varying the halo parameters in the range given above, and
matching the observed intensity to eq.~(\ref{phigamma}) to within 20\%, I
obtain the required WIMP annihilation cross section as a function of the WIMP
mass. This is the band marked ``halo $\gamma$'s'' in fig.~1.

The second band in fig.~1 comes from the requirement that the WIMP relic
density $\Omega h^2$ be in the cosmologically interesting range 0.025--1. ($h$
is the Hubble constant in units of 100km/s/Mpc.) The WIMP relic density is
related to the WIMP annihilation cross section through the approximate relation
(Kolb \& Turner 1990, Gondolo \& Gelmini 1991)
\begin{equation}
\label{omega}
\sigma v = \frac{ 2.0 \times 10^{-27} \rm{cm^3/s} }{ g g_{\star}^{1/2} x_f
  \Omega h^2 } ,
\end{equation}
where I have assumed that the annihilation cross section is dominated by the
$v=0$ term both in the galactic halo and at freeze-out (s-wave annihilation).
The freeze-out temperature $x_f m$ can be obtained solving $ x_f^{-1} +
\frac{1}{2} \ln(g_{\star}/x_f) = 80.4 + \ln(g m_\chi \sigma v) $, with $m_\chi$
in GeV and $\sigma v$ in cm$^3$/s. For a Majorana WIMP $g=2$, for a Dirac WIMP
$g=4$.  $g_{\star}$ is the effective number of relativistic degrees of freedom
at freeze-out: $g_{\star} \simeq 81$ before and $g_{\star} \simeq 16$ after the
QCD quark-hadron phase transition.  Letting $ \Omega h^2 $ vary in the
above range gives the relic density band in fig.~1.

The two bands intersect for WIMP masses between 1.2 and 50 GeV.  For example,
$\Omega = 0.2$ and $H = 60$ km/s/Mpc give the required cross section of $3
\times 10^{-26}$ cm$^3$/s at $m_\chi = 5$ GeV. (This case was presented in
Gondolo 1998a.)

\section{Constraints on candidates that couple to the Z boson.}

I assume in this section that $\chi\chi$ annihilation and $\chi$--nucleon
scattering are dominated by Z boson exchange. In this case, the annihilation
cross section reads
\begin{equation}
\sigma v = \frac{ G_F^2 } {\pi} \sum_f \beta_f \left[ 
m_f^2 \left( a_\chi^2 a_f^2 + v_\chi^2 v_f^2 - 2 v_\chi^2 a_f^2 \right) +
 2 m_\chi^2 v_\chi^2 \left( a_f^2 + v_f^2 \right) \right] ,
\end{equation}
where $ \beta_f = (1 - m_f^2/m_\chi^2 ) ^{1/2} $, $a_f=T_f$ and
$v_f=T_f-2e_f\sin^2\theta_W$ are the usual axial and vector couplings of
fermion $f$ to the Z boson, and $a_\chi$ and $v_\chi$ are the analogous
quantities for the $\chi$--Z coupling.

The $\chi$--nucleon scattering cross section
$\sigma_{\chi{\scriptscriptstyle\cal N}}$, which is limited by negative direct
dark matter searches, is related to the annihilation cross section by crossing
symmetry. The experimental bound on $\sigma_{\chi{\scriptscriptstyle\cal N}}$
depends on the WIMP mass and on the spin-dependent or spin-independent
character of the interaction (for a recent compilation of limits see Bernabei
et al.\ 1998). For Z boson exchange, the spin-dependent and spin-independent
$\chi$--nucleon cross sections read
\begin{equation}
  \sigma^{\rm SD}_{\chi {\scriptscriptstyle\cal N}} = \frac{ 6 \mu_{\chi{\scriptscriptstyle\cal N}}^2 }{ \pi} G_F^2
  a_\chi^2 \left[ \left( \Delta {\rm u} - \Delta {\rm d} \right)^2 + \left(
  \Delta {\rm s} \right)^2 \right] ,
\end{equation}
and
\begin{equation}
  \sigma^{\rm SI}_{\chi {\scriptscriptstyle\cal N}} = \frac{ \mu_{\chi{\scriptscriptstyle\cal N}}^2 }{ 4 \pi} G_F^2
  v_\chi^2 \left[ \left( 1 - 4 \sin^2\theta_W \right)^2 + 1 \right] .
\end{equation}
Here $\mu_{\chi{\scriptscriptstyle\cal N}} = m_\chi m_{\scriptscriptstyle\cal N}/(m_\chi+m_{\scriptscriptstyle\cal N}) $ is the
reduced $\chi$--nucleon mass, and $\Delta {\rm q}$ is the quark q contribution
to the spin of the proton. (From neutron and hyperon decay, $\Delta {\rm
  u}-\Delta {\rm d} = 1.2573\pm0.0028$, while $\Delta {\rm s}$ is uncertain:
$\Delta s=0$ in the naive quark model, $\Delta s=-0.11\pm0.03$ from deep
inelastic data, and $\Delta s=-0.15\pm0.09$ from elastic $\nu{\rm p}\to \nu{\rm
  p}$ data.)

The coupling of the $\chi$ to the Z boson also gives a contribution to the Z
boson decay width, if $m_\chi <  m_Z$. Namely,
\begin{equation}
  \Gamma(Z\to \chi\overline{\chi}) = \frac{ G_F m_Z } { 6 \sqrt{2} \pi }
  \left[ a_\chi^2 m_Z^2 \beta_\chi^3 + \beta_\chi v_\chi^2 \left(
    m_Z^2+2m_\chi^2 \right) \right] ,
\end{equation}
with $\beta_\chi = (1 - 4 m_\chi^2/m_Z^2 ) ^{1/2} $.  The experimental limit
is $\Gamma(Z\to\chi\overline{\chi}) < 5 $ MeV (Barnett et al.\ 1996).

Once a relation between $a_\chi$ and $v_\chi$ is specified, the experimental
bounds on $\sigma_{\chi{\scriptscriptstyle\cal N}}$ and
$\Gamma(Z\to\chi\overline{\chi})$ translate into a bound on $\sigma v$. Fig.~2
plots these bounds for two cases: a Dirac particle with $v_\chi=a_\chi$ ($V-A$
interaction), and a Majorana particle with $v_\chi=0$ (axial interaction).  For
Dirac particles, the interesting region is not fully excluded by these
constraints, but for Majorana particles it is. Hence the impact of these
constraints is model dependent.

\begin{figure}[htb]
\centerline{\epsfxsize = 0.8\hsize \epsfbox{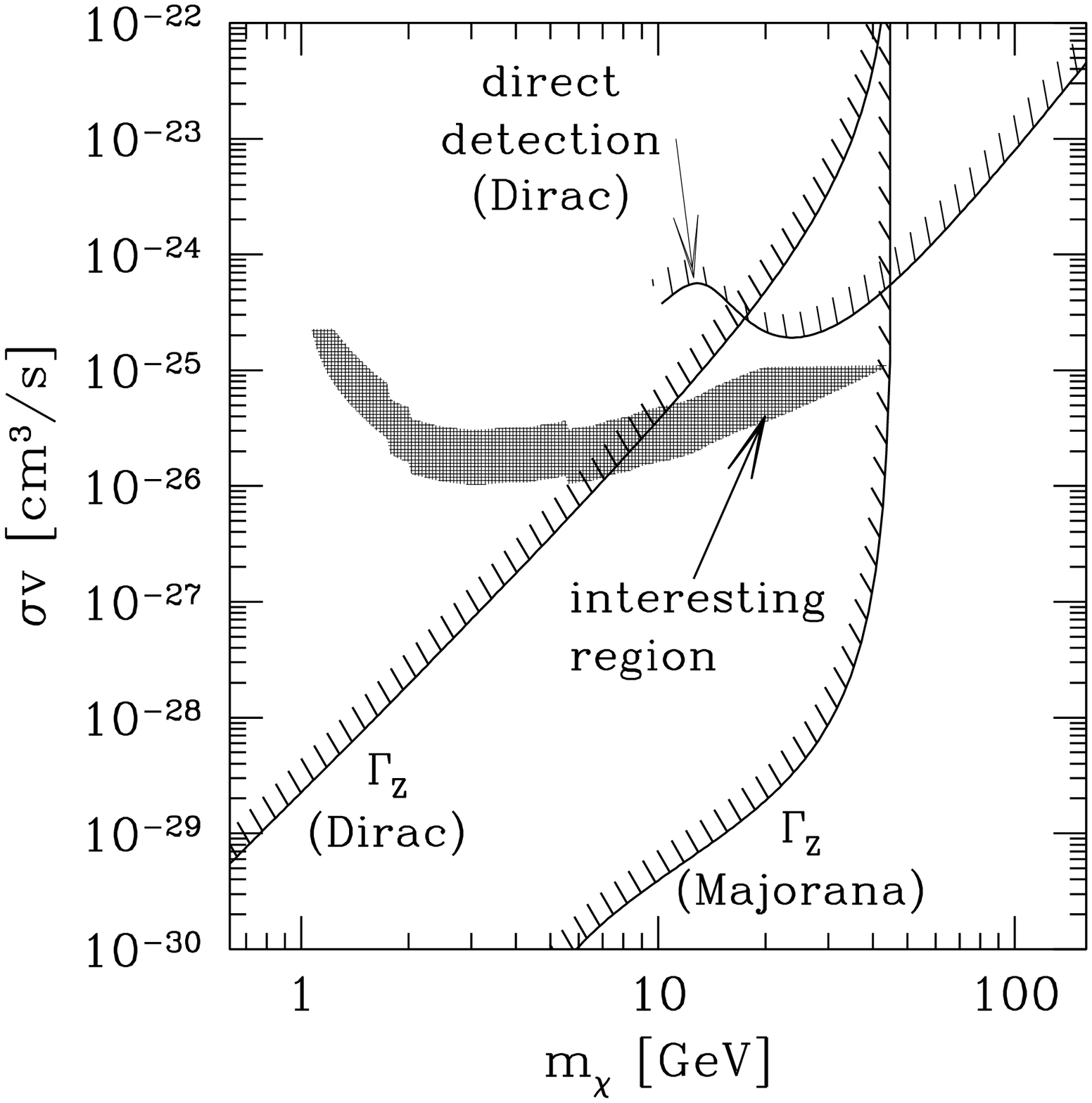}}
\fcaption{Constraints on the interesting region for a particle coupled to the Z boson.}
\label{fig:2}
\end{figure}

\section{Constraints from dark matter searches.}

A candidate WIMP has to satisfy constraints from negative dark matter searches.
In this section I consider indirect detection through production of rare cosmic
rays (antiprotons) and through neutrino production in the Sun and the Earth,
and direct detection through elastic scattering off nuclei in a laboratory
detector.

\begin{figure}[htb]
\centerline{\epsfxsize = 0.8\hsize \epsfbox{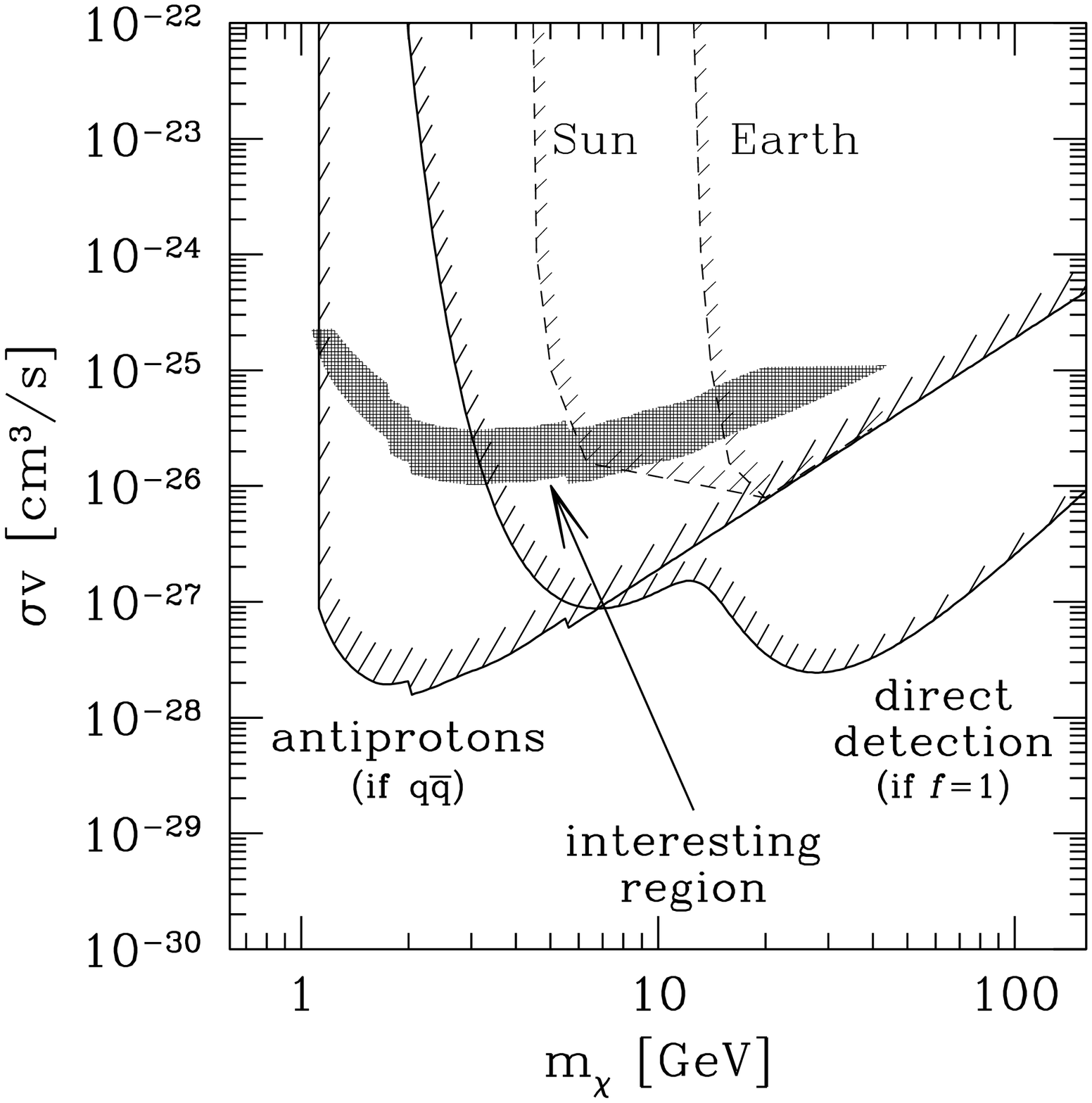}}
\fcaption{Constraints on the interesting region from dark matter searches.}
\label{fig:3}
\end{figure}

The most important constraint comes from the measured flux of cosmic ray
antiprotons. If the gamma-rays are produced in jets originated by quarks, there
is an associated production of antiprotons.  The ratio of antiproton and
gamma-ray fluxes is independent of the WIMP annihilation cross section and of
the local mass density, and the relative number of antiprotons and photons per
annihilation is fixed by the physics of jets. The antiproton flux at a ${\rm
  \bar{p}}$ kinetic energy $T$ at the top of the atmosphere is
\begin{equation}
  \phi_{\rm \bar{p}}(T) =  \frac{ d N_{\rm \bar{p}}}{ d T} \,
  \frac{ \sigma v }{ 4 \pi m_\chi^2 } \, \rho_{\rm
    loc}^2 v_{\rm \bar{p}} t_{\rm cont} \mu ,
\end{equation}
where $d N_{\rm \bar{p}}/ d T$ is the antiproton spectrum per annihilation,
$t_{\rm cont}$ is the $\bar{p}$ containment time, which in the diffusion model
of Chardonnet et al.\ (1996) is $t_{\rm cont} \simeq (1+p/3{\rm GeV})^{-0.6} 5
\times 10^{15} $ s, and $ \mu = [ T (T+2m_{\rm\bar{p}}) ] / [ (T+\Delta)
(T+\Delta+2m_{\rm\bar{p}}) ]$ takes into account solar modulation. Fig.~3 shows
the bound obtained by taking $\Delta=600$ MeV and imposing $\phi_{\rm
  \bar{p}}(150$--$300{\rm MeV}) < 3 \times 10^{-6} $ ${\rm
  \bar{p}}$/cm$^2$/s/sr/GeV (Moiseev et al.\ 1997). For WIMPs heavier than
$\sim 10$ GeV, a related constraint on the gamma-ray flux between 300 and 1000
MeV comes from the approximate relation $n_\gamma(\hbox{300--1000MeV}) \simeq
dN_{\bar{\rm p}}(\hbox{150--300MeV})/dT \times 10 {\rm GeV} $.  This gives
$\phi_\gamma(\hbox{300--1000MeV}) \lesssim 1.3\times 10^7$
pho\-tons/cm$^2$/s/sr if photons originate in quark jets with $m_\chi \gtrsim
10$ GeV.  Compared with the observed value of $\sim 8\times 10^7$
photons/cm$^2$/s/sr, it implies too low a gamma-ray flux below 1 GeV. However,
these bounds from antiprotons are uncertain because of uncertainties in the
antiproton propagation in the galaxy and in the effect of the solar modulation.
Moreover, these bounds depend sensitively on the relative branching ratios into
the various annihilation channels, and for example can be avoided if the
dominant decay channel is leptonic (see next section).

The $\chi$ fermions can also accumulate in the Sun and in the Earth, annihilate
therein and produce GeV neutrinos.  Accumulation is not efficient for WIMPs
lighter than the evaporation mass, which is $\sim 12$ GeV for the Earth and
$\sim 3$ GeV for the Sun. The curves marked ``Earth'' and ``Sun'' in fig.~3
show approximate constraints obtained from the experimental upper bounds on the
flux of through-going muons in the Baksan detector (Suvorova 1997).

Another limit comes from direct searches through the crossing
relation between annihilation and scattering cross sections. This relation can
be written as
\begin{equation}
  \mu_{\chi {\scriptscriptstyle\cal N}}^2 \sigma v = f m_\chi^2 c \sigma_{\chi
    {\scriptscriptstyle\cal N}} ,
\end{equation}
where $ c $ is the speed of light and $ f $ accounts for the details of the
interactions and of the nuclear structure.  The ``direct detection'' bound in
fig.~3 assumes $f=1$, but actually $f$ can range from 0 to infinity. For
example, consider a WIMP that couples to quarks and leptons only through
exchange of scalar bosons, and consider annihilation through s-channel exchange
and scattering through t-channel exchange.  For kinematical reasons, if the
scalar is CP-even, the annihilation cross section $\sigma v$ at $v=0$ vanishes
and the scattering cross section at small velocity is finite, so $f=0$. On the
other hand, if the scalar is CP-odd, it is the scattering cross section that
vanishes and the annihilation cross section is finite, so $f$ is infinite.
Choosing the relative strength of CP-odd and CP-even exchange, the magnitude of
the annihilation and scattering cross sections can be tuned. An explicit
example is given in the next section.

\section{A candidate WIMP can be a sterile neutrino.}

The analysis in the previous paragraphs may have been
instructive, but the conclusions are so model-dependent that the discussion 
would be clearer if done in a specific model. Here I show
that a particle with the required properties is a sterile right--handed
neutrino $\nu_s$ in a model with an extended Higgs sector.

The Higgs sector contains two Higgs doublets $H_1$ and $H_2$ and a Higgs
singlet $N$. I assume the following Higgs potential
\begin{eqnarray}
\label{eq:V}
  V_{\rm Higgs} = && 
  \lambda_1 \left( H_1^\dagger H_1 - v_1^2 \right)^2 +
  \lambda_2 \left( H_2^\dagger H_2 - v_2^2 \right)^2 +
  \lambda_3 \left| N^2 - v_N^2 \right|^2 + \nonumber\\&&
  \lambda_4 \left| H_1 H_2 - v_1 v_2 \right|^2 + 
  \lambda_5 \left| H_1 H_2 - v_1 v_2 + N^2 - v_N^2 \right|^2 +
  \lambda_6 \left| H_1^\dagger H_2 \right|^2 .
\end{eqnarray}
To fix the notation, $ H_1 H_2 = H_1^0 H_2^0 - H_1^- H_2^+ $.
Taking all the $\lambda$'s real and positive guarantees that the absolute
minimum of the Higgs potential is at $ \langle H_1^0 \rangle = v_1$, $ \langle
H_2^0 \rangle = v_2 $, and $ \langle N \rangle = v_N $, with the vacuum
expectation values of all the other fields vanishing.

Through the Yukawa terms
\begin{equation}
  {\cal L}_{\rm Yukawa} = f_d Q H_1 d_R + f_u Q H_2 u_R + f_e L H_1 e_R + h
  \overline{\nu_{sR}^c} \nu_{sR}^{\phantom{c}}\relax N,
\end{equation}
$H_1$ gives masses to the up--type quarks and the charged leptons, $H_2$ gives
masses to the down--type quarks, and $N$ gives a Majorana mass to the
right--handed neutrino. There is no mixing of the right--handed neutrino with
ordinary neutrinos, otherwise the new neutrino would have decayed in the early
universe and would not be in the galactic halo at present.\footnote{F.~Vissani
  has kindly pointed out that a particle that does not mix with ordinary
  neutrinos should not be called a ``neutrino.'' I keep this name for lack of a
  better one.}

There are 2 would-be Goldstone bosons -- a charged one $G^\pm$ and a neutral one
$G^0$ -- and 6 physical Higgs bosons: a charged one $H^\pm$, three neutral
``scalars'' $S_i$ ($i=1,\dots,3$), and two neutral ``pseudoscalars'' $P_i$
($i=1,2$).

The physical charged Higgs field $ H^\pm = H^{\mp *} \sin\beta + H_2^\pm
\cos\beta $ has squared mass
\begin{equation}
  m_{H^+}^2 = \lambda_6 v^2 .
\end{equation}
As usual, $v = \sqrt{v_1^2 + v_2^2} $ and $\tan\beta = v_2/v_1$.

The squared mass matrix of the two physical pseudoscalar neutral Higgs fields
reads 
\begin{equation}
  {\cal M}_P^2 = 
  \left( \begin{array}{cc}
    (\lambda_4 + \lambda_5) v^2 &
     2 \lambda_5 v v_N \\
    2 \lambda_5 v v_N &
     4 (\lambda_3 + \lambda_4) v_N^2
  \end{array} \right)
\end{equation}
in the basis $ \sqrt{2} ( \sin\beta {\rm Im} H_1^0 + \cos\beta {\rm Im} H_2^0 ,
{\rm Im} N )$. The orthogonal combination of fields is the would-be Goldstone
boson $G^0$. Let $U^P$ be the unitary matrix that diagonalizes ${\cal M}_P^2$,
namely $ P_i = U^P_{i1} A + U^P_{i2} N_I$ where $i=1,2$.

The scalar Higgs boson mass matrix is 
\begin{equation}
  {\cal M}_S^2 = 
  \left( \begin{array}{ccc}
    4 \lambda_1 v_1^2 + (\lambda_4 + \lambda_5) v_2^2 &
     (\lambda_4 + \lambda_5 ) v_1 v_2 & 
      2 \lambda_5 v_2 v_N \\
    (\lambda_4 + \lambda_5 ) v_1 v_2 &
     4 \lambda_2 v_2^2 + (\lambda_4 + \lambda_5) v_1^2 &
      2 \lambda_5 v_1 v_N \\
    2 \lambda_5 v_2 v_N &
     2 \lambda_5 v_1 v_N &
      4 (\lambda_3 + \lambda_5) v_N^2
  \end{array} \right) 
\end{equation}
in the basis $ \sqrt{2} ( {\rm Re} H_1^0, {\rm Re} H_2^0, {\rm Re} N ) $. Its
mass eigenstates $S_i$ ($i=1,\dots,3$) are obtained as $S_i = \sqrt{2} (
U^S_{i1} {\rm Re} H_1^0 + U^S_{i2} {\rm Re} H_2^0 + U^S_{i3} {\rm Re} N )$.

The original Higgs fields can be expressed in terms of the physical fields as
\begin{eqnarray}
  H_1^0 &=& v_1 + \sqrt{\textstyle{1\over2}} 
    \left( U^S_{i1} S_i + i U^P_{i1} P_i \sin\beta
  \right) , \\
  H_2^0 &=& v_2 + \sqrt{\textstyle{1\over2}} 
    \left( U^S_{i2} S_i + i U^P_{i1} P_i \cos\beta
  \right) , \\
  N &=& v_N + \sqrt{\textstyle{1\over2}} 
    \left( U^S_{i3} S_i + i U^P_{i2} P_i \right) .
\end{eqnarray}
This gives the interactions of the Higgs bosons with the quarks, the leptons
and the right--handed neutrino,
\begin{eqnarray}
  {\cal L}_{\rm int} &=& \frac{g m_{\nu_s}}{2m_W} \frac{v}{v_N} \left[ 
  U^S_{i3} S_i \overline{\nu}_s \nu_s + i
  U^P_{i2} P_i \overline{\nu}_s \gamma_5 \nu_s \right] \\ &+&
\frac{g m_u}{2m_W} \left[ 
  \frac{U^S_{i2}}{\sin\beta} S_i \overline{u} u + i
  \cot\beta U^P_{i1} P_i \overline{u} \gamma_5 u \right] \\ &+&
\frac{g m_d}{2m_W} \left[ 
  \frac{U^S_{i1}}{\cos\beta} S_i \overline{d} d + i
  \tan\beta U^P_{i1} P_i \overline{d} \gamma_5 d \right]
\end{eqnarray}

It is then easy to work out the annihilation cross section,
\begin{equation}
  \sigma v = \frac{ G_F^2 m_\nu^4} {\pi} \, \frac{v^2}{v_N^2} \,
  \left[ \sum_{k=1}^2  \frac{ U^P_{k2} U^P_{k1} }{ m_{P_k}^2 - 4 m_\nu^2 } 
     \right]^2 \,
   \sum_f c_f \beta_f m_f^2 \kappa_f^2 ,
\end{equation}
and the scattering cross section off nucleons,
\begin{equation}
  \sigma_{\nu{\scriptscriptstyle\cal N}} = \frac{G_F^2}{\pi} \, \frac{2
    m_{\scriptscriptstyle\cal N}^4 m_\nu^4}{\left( m_{\scriptscriptstyle\cal N}
    + m_\nu \right)^2} \, \frac{v^2}{v_N^2} \, \left[ \sum_{j=1}^3
  \frac{U^S_{j3}}{m_{S_j}^2} \left( \frac{k_d U^S_{j1}}{\cos\beta} + \frac{ k_u
    U^S_{j2}}{\sin\beta} \right) \right]^2 .
\end{equation}
Here $c_f=3$ for quarks, $c_f=1$ for leptons, $\kappa_f=\cot\beta$ for up--type
quarks, and $\kappa_f=\tan\beta$ for down--type quarks and leptons. Moreover, $
k_d = \langle m_d \overline{d} d + m_s \overline{s} s + m_b \overline{b} b
\rangle = 0.21$ and $ k_u = \langle m_u \overline{u} u + m_c \overline{c} c +
m_t \overline{t} t \rangle = 0.15$.

Table 1 lists important quantities for two interesting cases: a 3 GeV neutrino
and a 7 GeV neutrino. 

\begin{table}[h]
\tcaption{Two interesting cases.}\label{tab:1}
\small
\begin{tabular}{||c|c|c|c||}\hline\hline
 {} & ``3 GeV'' & ``7 GeV'' & \begin{minipage}{1in} \begin{center}
 Experimental/ Observational Values \end{center} \end{minipage} \\
\hline
$m_\nu$ [GeV] & $3$ & $7$ & {} \\
$\phi_\gamma(56^\circ)$ {[photons($>$1GeV)/cm$^2$/s/sr]}
& $7.3\times 10^{-7}$ & $7.4 \times 10^{-7}$  & $\sim 7.6 \times 10^{-7}$  \\
$\Omega$ ~~($H=60$ km/s/Mpc) & $0.11$ & $0.12$ & $\sim 0.2$ \\
$\phi_{\rm\bar{p}} (200{\rm MeV})$ {[$\overline{\rm p}$/cm$^2$/s/sr/GeV]} &
$\sim 0$ & $7\times 10^{-6}$ & $\lesssim 3\times 10^{-6} $ \\
$\sigma_{\chi{\scriptscriptstyle\cal N}}$ [pb] & 
$1.2\times 10^{-3}$ & $1.3 \times 10^{-4}$ & see text \\
$m_{P_1}, m_{P_2}$ [GeV] & $5,155$  & $4.2,156$ & {} \\
$m_{S_1}, m_{S_2}, m_{S_3}$ [GeV] & $5,155,220$ & $4.2,156,220$ & {} \\
$m_{H^\pm}$ [GeV] & $110$ & $110$ & $\gtrsim 53$ \\
$\Gamma(Z\to f\overline{f}P,S)/\Gamma(Z\to f\overline{f})$ & $1.3\times
10^{-6}$ & $6.3\times 10^{-6}$ & $\lesssim 10^{-4}$ \\
``$\sin^2(\beta-\alpha)$'' (Zh) & $2.3\times 10^{-7}$ & $2.3\times 10^{-6}$ &
$\lesssim 10^{-2}$ \\
``$\cos^2(\beta-\alpha)$'' (hA) & $6 \times 10^{-7} $ & $6\times 10^{-6} $ &
$\lesssim 0.3$ \\
\hline\hline
\end{tabular} 
\end{table}

For the ``1 GeV'' model, the parameters are $h=0.9$, $v/v_N=52$,
$\tan\beta=v_2/v_1=40$, and $\lambda_1,\ldots,\lambda_6=g^2$ (the square of the
SU(2) coupling constant). The gamma-ray intensity is close to the observed one
and the relic density is in the cosmologically interesting range. There is
practically no antiproton flux from annihilations because only the
$\tau^+\tau^-$ channel is effective. The scattering cross section off nucleons
is compatible with the present bound of $2\times 10^{-2}$ pb. Constraints on
the Higgs sector are discussed below.

For the ``7 GeV'' model, the parameters are $h=0.8$, $v/v_N=20$,
$\tan\beta=33$, and $\lambda_1,\ldots,\lambda_6=g^2$.  The gamma-ray intensity
is closed to the observed one, and the relic density is in the cosmologically
interesting range. The antiproton flux is slightly higher than the measured
one, but, given the big uncertainties in its estimation, it is compatible with
it. The scattering cross section is smaller than the present limit of $3\times
10^{-4}$ pb. The Higgs sector is discussed in the following.

Two Higgs bosons ($S_1$ and $P_1$) are light and one has to worry about their
production at colliders. But this production is very suppressed because the
light Higgs bosons are mostly singlets.  First, the LEP bounds on the search of
Higgs bosons in two-doublet models are satisfied.  Since there is an additional
Higgs singlet, $\sin^2(\beta-\alpha)$ in the bound from $e^+ e^- \to Z h$ must
be replaced with $| U^S_{11} \cos\beta + U^S_{12} \sin \beta |^2$, and
$\cos^2(\beta-\alpha)$ in the bound from $e^+ e^- \to hA$ must be replaced with
$ | U^P_{12} U^S_{12} - U^P_{11} U^S_{11} |^2$. In the allowed region, the
reinterpreted $\sin^2(\beta-\alpha)$ and $\cos^2(\beta-\alpha)$ are smaller
than $5 \times 10^{-5}$, and are not excluded by accelerator searches (Decamp
et al.\ 1992).  Secondly, Higgs bremsstrahlung from final state leptons and
quarks in Z decays is suppressed by the small mixing between singlet and
doublet Higgs bosons.  In the present model,
\begin{equation}
\label{brems}
\frac{ \Gamma\!(Z\to f\overline{f}X) }{ \Gamma\!(Z\to f\overline{f}) }
= \frac{ \sqrt{2} G_F } { 4 \pi^2} \, g\!\left(\frac{m_X}{m_Z}\right) \,
c_f \, m_f^2 A_{Xf}^2 ,
\end{equation}
where $A_{P_if} = U^P_{i1} \cot\beta$, $A_{S_if} = U^S_{i1}/\cos\beta$ for
up--type quarks, $A_{P_if} = U^P_{i1} \tan\beta$, $A_{S_if} =
U^S_{i2}/\sin\beta$ for down--type quarks and leptons, $c_f = 3$ for quarks and
$c_f=1$ for leptons. The function $g(y)$ comes from the phase-space
integration, and is $g(y) \cong 1$ at $m_X=12$ GeV. The values obtained in the
two examples (see table 1) are lower than the LEP constraint on two leptons +
two jets production $ \Gamma\!(Z\to f\overline{f}X) / \Gamma\!(Z\to
f\overline{f}) < 1.5 \times 10^{-4} $ at $m_X = 12 $ GeV (Decamp et al.\ 1992).

I conclude that these two examples satisfy the experimental and observational
constraints considered. A more general scan of the model parameter space is
under way (Gondolo 1998b).

\section{Conclusions}

I have shown that it may be possible to explain the Dixon et al.\ gamma-ray
emission from the galactic halo as due to halo WIMP annihilations. Not only the
intensity and spatial pattern of the halo emission can be matched but also the
relic density of the candidate WIMP can be in the cosmologically interesting
domain.  After a model-independent analysis to learn about the properties of a
suitable candidate, I have presented a working model: a sterile neutrino in a
model with an extended Higgs sector. Two examples in parameter space indicate
the existence of an interesting region in the model parameter space in which
present observational and experimental constraints are satisfied and the
gamma-ray emission is reproduced.

\section{Acknowledgments}

I thank Georg Raffelt and Bernd Kniehl for the invitation to give a talk at
this Workshop, Apostolos Pilaftsis and Ara Ioannisian for a discussion on
constraints on models with extended Higgs sectors, and Leo Stodolsky for
pointing out Higgs bremsstrahlung.

\section{References}

\noindent Barnett R. M. et al.\ (Particle Data Group) 1996, {\it Phys. Rev.}
{\bf D54}, 1.

\noindent Bergstr\"om L., Edsj\"o J., and Ullio P. 1998, astro-ph/9804050
(April 1998).

\noindent Bernabei R. et al.\ 1998, University of Rome preprint ROM2F/98/08
(February 1998).

\noindent Chardonnet P., Mignola G., Salati P., and Taillet R. 1996, {\it
  Phys. Lett.} {\bf B384}, 161.

\noindent Decamp D. et al.\ 1992, {\it Phys. Rep.} {\bf 216}, 253.

\noindent Dixon D. D. et al. 1998, in {\it Sources and
  Detection of Dark Matter}, 
Marina del Rey, 

California, February 1998; also astro-ph/9803237 (March 1998).

\noindent Gondolo P. and Gelmini G. 1991, {\it Nucl. Phys.} {\bf B360}, 145.

\noindent Gondolo P. 1998a, comment to Dixon's talk in {\it Sources and
  Detection of Dark Matter}, 

Marina del Rey, California, February 1998 (to
appear in the Proceedings).

\noindent Gondolo P. 1998b, in preparation.

\noindent Gunn J. E. et al.\ 1978, Ap. J. {\bf 223}, 1015.

\noindent Kolb E. W. and Turner M. S. 1990, {\it The Early
    Universe} (Addison-Wesley, Redwood 

City).

\noindent Moiseev A. et al.\ (BESS Collab.), 1997, Ap. J. {\bf 474}, 479.

\noindent Turner M. 1986, {\it Phys. Rev.} {\bf D34}, 1921.

\end{document}